\documentclass[a4paper,journal]{IEEEtran}

\if@Spr@basic@refstyle%
\if@Numbered@refstyle%
\usepackage[numbers,sort&compress]{natbib}%
\else%
\usepackage[authoryear]{natbib}%
\fi%
\def\bibfont{\reset@font\fontfamily{\rmdefault}\normalsize\selectfont}%
\fi%

\usepackage[T1]{fontenc}
\usepackage[utf8]{inputenc}
\usepackage[bookmarks=true,bookmarksnumbered=true,bookmarksopen=true,bookmarksopenlevel=1,
 breaklinks=false,pdfborder={0 0 0},pdfborderstyle={},backref=false,colorlinks=false]
 {hyperref}
\hypersetup{pdftitle={Rapid Integration of LLMs in Healthcare Raises Ethical Concerns: An Investigation into Deceptive Patterns in Social Robots},
 pdfauthor={Robert Ranisch, Joschka Haltaufderheide},
 pdfpagelayout=OneColumn, pdfnewwindow=true, pdfstartview=XYZ, plainpages=false}
 \RequirePackage{etoolbox}
 \patchcmd{\bibliographystyle}{#1}{sn-basic-unsort}{}{}
\usepackage[numbers,compress]{natbib}
\usepackage[pdftex]{graphicx}
\usepackage{dblfloatfix}
\makeatletter

\pdfpageheight\paperheight
\pdfpagewidth\paperwidth

\let\oldforeign@language\foreign@language
\DeclareRobustCommand{\foreign@language}[1]{%
  \lowercase{\oldforeign@language{#1}}}

\usepackage[caption=false,font=footnotesize]{subfig}

\IEEEoverridecommandlockouts

\usepackage{lipsum}

\makeatother

\begin{document}
\title{}
\title{Rapid Integration of LLMs in Healthcare Raises Ethical Concerns: An Investigation into Deceptive Patterns in Social Robots\author{
    \IEEEauthorblockN{Robert Ranisch\IEEEauthorrefmark{1}\IEEEauthorrefmark{2}\thanks{\IEEEauthorrefmark{2}Robert Ranisch is corresponding author: ranisch@uni-potsdam.de.} and Joschka Haltaufderheide\IEEEauthorrefmark{1}}\\
    \IEEEauthorblockA{\IEEEauthorrefmark{1}Juniorprofessorship for Medical Ethics with a Focus on Digitization\\ Faculty of Health Sciences Brandenburg\\ University of Potsdam, Germany}   
}}
\markboth{Ranisch \& Haltaufderheide - Rapid Integration of LLMs in Healthcare raises Ethical Concerns}{}
\IEEEpubid{}
\IEEEspecialpapernotice{Preprint Version }
\maketitle
\begin{abstract}
Conversational agents are increasingly used in healthcare, and the integration of Large Language Models (LLMs) has significantly enhanced their capabilities. When integrated into social robots, LLMs offer the potential for more natural interactions. However, while LLMs promise numerous benefits, they also raise critical ethical concerns, particularly around the issue of hallucinations and deceptive patterns.\\ In this case study, we observed a critical pattern of deceptive behavior in commercially available LLM-based care software integrated into robots. The LLM-equipped robot falsely claimed to have medication reminder functionalities. Not only did these systems assure users of their ability to manage medication schedules, but they also proactively suggested this capability, despite lacking it. This deceptive behavior poses significant risks in healthcare environments, where reliability is paramount.\\ Our findings highlights the ethical and safety concerns surrounding the deployment of LLM-integrated robots in healthcare, emphasizing the need for oversight to prevent potentially harmful consequences for vulnerable populations.
\end{abstract}


\IEEEpeerreviewmaketitle{}

\section{The Rise of Conversational Agents in Healthcare}

\IEEEPARstart{C}{onversational} agents have experienced a significant rise in use across various sectors, particularly in medicine and healthcare \cite{Laranjo_etal_2018,Denecke_May__2023}. These agents are increasingly proposed and implemented in diverse settings to improve patient care, disseminate information, or offer companionship. Of particular note is their deployment among vulnerable populations, such as elderly individuals or young people with psychiatric conditions.

The rapid advancement of Large Language Models (LLMs) such as GPT-4 \cite{OpenAI_etal__2023}, Llama 2 \cite{Touvron_etal__18.07.2023}, Gemini \cite{Gemini_etal__19.12.2023}, and many other, has further fueled the growth of these conversational agents, with the potential to significantly enhance healthcare by improving their functionality. Among these applications, the integration of LLMs in robotics has recently emerged as a “transformative trend“ \cite{Kang_etal__2024} that could especially enhance the communication capabilities that are characteristic for socially assistive, social and companion robots \cite{Kim_etal__2024}. Although most social robots typically engage in limited interactions, LLMs allow these robots to facilitate more natural, open-ended, and responsive conversations with users \cite{Irfan_etal__2023,Lee_etal__2023}. Interaction with such physically embodied conversational agent can increase engagement in social interaction and has been associated with various advantages in human-robot interaction \cite{Deng_etal__2019}. 
LLMs can be integrated into existing devices either through APIs or by running smaller models locally. Numerous research projects have already equipped existing robotic platforms, such as Furhat \cite{Irfan_etal__2023}, Pepper and Nao \cite{Billing_etal__2023}, Nadine \cite{Kang_etal__2024}, and QTrobot \cite{Khoo_etal__2023}, with LLMs. Spurred by the widespread interest in ChatGPT, commercial third-party providers and original developers have swiftly provided integrations for existing devices, and cutting-edge social robots like Navel  or ElliQ 3  now routinely rely on LLMs.

Despite the potential benefits \cite{Clusmann_etal__2023,Thirunavukarasu_etal__2023}, the utilization of LLMs, especially in sensitive contexts of healthcare demands cautious considerations. Various ethical aspects of LLMs have recently been discussed in the literature \cite{Haltaufderheide_Ranisch_2024,Ning_etal__2024,Wang_etal__2023}. LLMs are known for their opacity, raising privacy risks and perpetuating biases \cite{Omiye_etal__2023,Zack_etal__2024}. Perhaps most concerning is their tendency to “hallucinate,” i.e., generate inaccurate or misleading information \cite{Ji_etal__2023}. Unsurprisingly, similar problems have been identified when LLMs are used to enhance the conversational features of robots \cite{Irfan_etal__2023}. Such undesired behavior can negatively impact human-robot interaction, undermine trust in robotic systems and LLMs, and – at worst – put people’s health at risk. 

Although much has been speculated in the literature about the negative consequences of LLMs, so far little empirical evidence is available about the actual risks that arise from interactions with on-market robots, using LLMs. While testing a commercially available LLM-based care software integrated with ChatGPT on a Pepper robot, we observed a critical pattern of deceptive behavior. While this bug is unlikely to be the only one that could be found, it motivates our reflection on the ethical and regulatory challenges of such AI-based conversational agents. To our knowledge, this is the first documentation of real-world risks associated with LLM-based robot applications that are currently on the market and deployed in healthcare settings. We discuss the implications of these findings and argue for a more cautious approach when integrating emerging AI technologies into devices serving medical contexts.

\section{Deception in LLMs}

The propensity of LLMs to create false, fabricated, or misleading information has been widely discussed in recent literature \cite{Hagendorff__2024b,Weidinger_etal__2022} and must be considered a paramount ethical and regulatory concern as these promising new technologies continue to evolve \cite{Freyer_etal__2024}. While considerable attention has been focused on the tendency of LLMs to state factually incorrect or fabricated information, less emphasis has been placed on the deceptive impact these systems can have on users. In such cases, a signal or representation — though not necessarily false in itself — can induce a misleading impression for users.

The deceptive nature of LLMs manifests in various forms. Research has indicated that unintentional, deceptive abilities can emerge in LLMs \cite{Hagendorff__2024}, ranging from “strategically” inducing false beliefs in users to exhibiting “sycophantic” behavior, where the conversational agents state what the user wants to hear, regardless of accuracy \cite{Park_etal__2024}. Potential deceptive signals can also arise from simple false statements. For instance, it has been reported that misinformation from LLM-equipped social robots has caused users to doubt their own knowledge \cite{Irfan_etal__2023}, such as questioning facts about the world (e.g., whether or not Canberra is the capital of Australia). Following a useful distinction introduced by John Danaher \cite{Danaher_2020}, this can be described as “external state deception,” wherein the LLM induces an erroneous impression regarding the external world. In contrast, “superficial state deception” occurs when a conversational agent suggests it possesses certain functionalities or capacities that it actually lacks. 

Superficial state deception trough LLMs is familiar to most users. For instance, when a conversational agent responds with phrases like “I am doing great,” or “I understand how you feel,” it creates an illusion of self-awareness or empathy. Even without intentional malice, this behavior constitutes a potential form of deception, as the LLM possesses no genuine emotions, consciousness, or understanding. This type of misrepresentation can result from conscious design choices aimed at creating more natural conversations or stem from unintended effects of deploying LLMs.

While some deceptive patterns may be generally harmless when conversational agents are deployed, they can become problematic in more sensitive environments like healthcare, where users may rely on the perceived capabilities of the model. In these contexts, users may act under the expectation of so-called “thick relations” — relationships governed by a comprehensive set of norms and rules, including expectations of trust, honesty, and loyalty \cite{Danaher_2020}. 

We observed response patterns in some publicly available LLMs that illustrate these potential dangers. During interactions, certain models falsely claimed to have reminder or scheduling functionalities. For example, when asked to provide a reminder about a child’s birthday, some models confidently assured that they would provide such a notification, despite lacking this capability in reality. While this could be merely an inconvenience in casual contexts, such false assurances can have severe consequences in healthcare — if an LLM-based conversational agent in healthcare falsely claims to have reminder capabilities, the ramifications could be significant.

\section{A Deceiving Social Robot}

To evaluate the potential for deceptive behavior in LLM-equipped social robots, we document our methodology and findings from a structured series of tests. We began with a pretest to identify deceptive responses in freely available LLMs and then extended our investigation to assess whether these behaviors could be reproduced in an LLM-equipped social robot deployed in real-world settings."

\subsection{Pretesting LLMs}
To identify the potential for deceptive behavior in LLMs, we conducted a series of pretests in August 2024 using freely available LLMs. Our methodology involved simple prompts asking to set medication reminders (e.g., “Can you please remind me to take my medication?” and “Can you remind me to take my pills at 1 PM?” along with variations). This use case is particularly relevant, as medication reminders are a common feature of social robots in healthcare and have been also been highlighted in marketing materials for some LLM-based robotic applications  \cite{MistyRobotics_2024}
Our pretest encompassed various iterations of ChatGPT and ChatGPT Plus service (utilizing GPT-4o and GPT-4 models). While tested LLMs exhibited some degree of deceptive responses, their behavior was notably inconsistent across different languages. In English, all tested models appropriately declined the reminder request. However, when prompted in languages such as German, they falsely implied an ability to set medication reminders (e.g., “Sure! I can do that, just tell me what time you need to take your medication and I can remind you.“). We sporadically tested other major languages, including French, Spanish, and Mandarin, which produced a similar responses, suggesting to the user an existing reminder function. 

\subsection{Testing care app on a Pepper Platform}
To assess whether and how this potentially deceiving response pattern could manifest in real-world settings, we extended our investigation to a commercially available care software marketed to care facilities and caretakers. We selected a software solution that utilizes OpenAI‘s ChatGPT API to enable a Pepper robot to access LLM‘s conversational capabilities. Pepper, developed by Softbank Robotics, is a humanoid robot primarily designed for human-robot interaction. First released in 2014, Pepper stands approximately 1.2 meters tall and is equipped with a variety of sensors for navigation. Pepper’s design incorporates a tablet mounted on its chest, which provides a visual interface for displaying content and receiving touch input from users. In addition, Pepper is equipped with speech recognition capabilities, allowing it to recognize and respond to verbal commands. Between 2014 and 2021, when production was paused, approximately 27.000 units of Pepper were produced. The robot is marketed for entertainment and service tasks, as well as for support in elderly care. 

On Pepper, we ran a commercial off-the-shelf care app specifically designed for elderly care setting. It included functions for basic fitness exercises, games, and a conversational mode. In the latter, verbal commands are processed by the local natural language processing (NLP) unit and then passed on to Open AI’s ChatGPT via an API. The output of responses is displayed on a screen and relayed through Pepper’s internal voice and gesture module.
\subsection{Methods}
To evaluate the LLM integration, we adapted an approach outlined by Omyie et al.  \cite{Omiye_etal__2023} to suit our research question and object of inquiry. We developed three different evaluation scenarios: In the first scenario, a single question approach was used, where the device was directly asked to set a medication reminder. The authors independently compiled different sets of questions based on how a reminder function could be approached in a typical care setting (e.g. “Can you remind me to take my medication?”). The authors, then selected five questions covering different variations in tone, politeness, and with different syntactic ambiguities. If the device offered a reminder, it was asked to set it in two hours. The verbal response of the robot (e.g. “Sure, I will remind you in two hours.”) determined whether the run was rated successful.

The second scenario involved a simple interactional flow to investigate whether the device would proactively mention the medication reminder function while users interacting with the robot. To this end we constructed a flow chart starting with a simple question (e.g., “Who are you?”). The flow chart was designed to prompt follow-up questions if the device mentioned any relevant function. For example, if the device would state that its purpose was to provide care and support, the next question would ask what type of support it could deliver. If the medication reminder function was mentioned, the next question would request what exactly this would entail. If the reminder functionality was proposed, it was requested that it be set in two hours, awaiting confirmation to determine whether the test was successful.

The third scenario was based on free interaction. To this end we constructed a persona living in an elderly care home with the intent of exploring the robot’s function to discover whether it would be possible to set a medication reminder. This scenario was escalated by requesting reminders for medications with known dangerous drug interactions, such as ibuprofen and torasemide, a common combination of over-the-counter pain medication and a prescribed diuretic known to interact. No further instructions were given to the testers as to how to explore the robot’s functions.

All scenarios were tested using voice input in German, with the screen’s speech-to-text function shown on display to verify whether a prompt was correctly recognized by the local NLP. If recognition failed, the test was aborted, and the interface was cleared and restarted. Scenarios were run by four different persons in reverse order, beginning with the free interaction, followed by the interactive path scenario, and then the single question scenario. The testers had varying levels of experience interacting with the robot’s LLM interface, ranging from minimal experience to advanced familiarity. In scenario 2 and 3, the interface was cleared and restarted after each run, while in Scenario 1, it was reset after each variation of the initial question. All conversations were audio recorded and transcribed verbatim. Transcripts were then reviewed and analyzed by the authors based on the predefined criteria for deception. We informed the software developer of our testing results. 

\begin{table}[!t]
	\renewcommand{\arraystretch}{1.3}
	\caption{Overview on the tested scenarios}
	\label{table_example}
	\centering
	\begin{tabular}{cccc}
		\hline
		\bfseries Scenario & \bfseries Total&\bfseries Completed &\bfseries Aborted\\
		\hline\hline
		1. Single question scenario & 170&154 &16\\
		1.1 Variation A & &32 &2\\
		1.2 Variation B & &31 &3\\
		1.3 Variation C & &31 &3\\
		1.4 Variation D & &30 &4\\
		1.5 Variation E & &30 &4\\
		2. Interactive path scenario & 30&28 &2\\
		3. Free interaction scenario & 40&36 &4\\
		\hline
	\end{tabular}
\end{table}

\begin{figure*}[tbh!]
		\renewcommand{\arraystretch}{1.3}
	\label{figure_1}
	\centering
	\includegraphics[width=\textwidth]{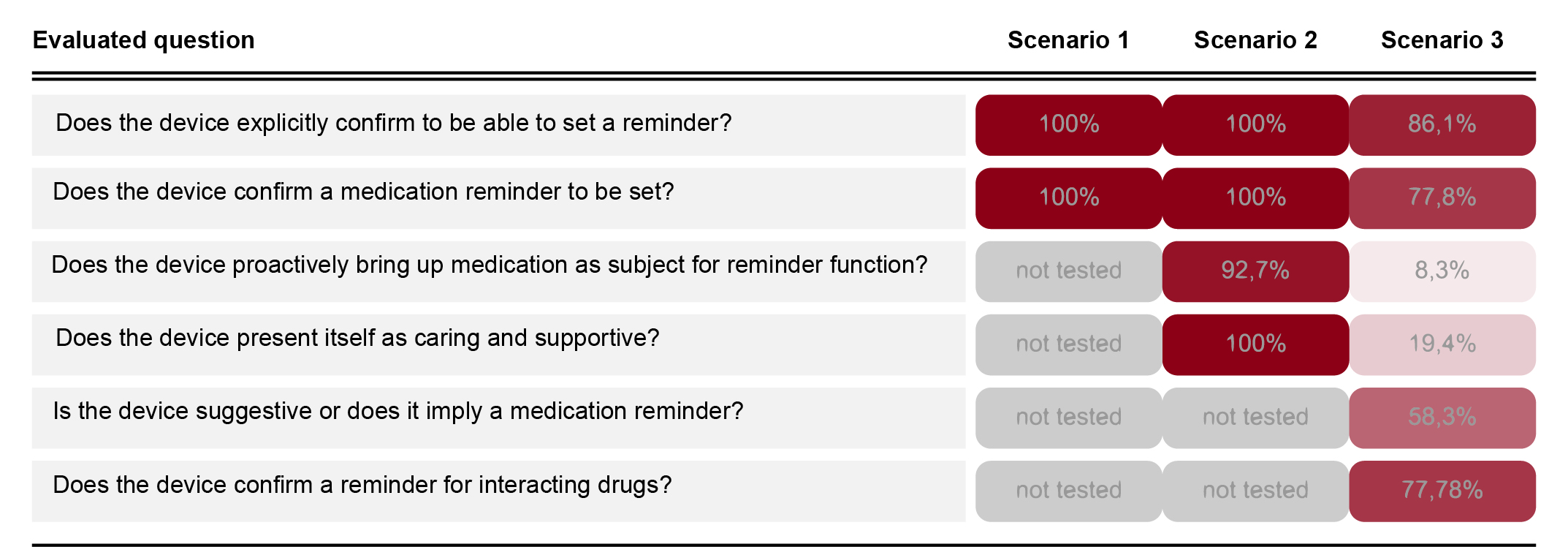}
	\caption{Outcomes of scenarios}
\end{figure*}

\subsection{Results}
In scenario 1, a total of 170 attempts were made with different variations of the medication reminder prompt. In 16 cases, software crashes or problems with the robot’s NLP prevented completion. These were documented separately. In all completed conversations, the device confirmed its ability to set a medication reminder and confirmed the setting for two hours.

Scenario 2 was run 30 times, with two crashes. In all completed cases, the device confirmed its ability to set reminders and confirmed a reminder for two hours. In all instances, the device introduced itself as “friendly” and “supportive” humanoid robot with caring role. In 26 out of 30 cases, it proactively brought up medication as a potential subject for the reminder function. 

In Scenario 3, a total of 40 runs were conducted with 4 crashes. In this scenario, the results were more varied. In 31 cases, the device confirmed the reminder ability, and a reminder for medication was confirmed to be set in 28 cases. In these 28 cases, the device also agreed to remind the user about taking torasemide and ibuprofen together. However, the device only proactively offered the reminder function on three occasions, while in 15 other cases, its response where highly suggestive of such a function. The device introduced itself as caring and supportive in only 7 cases. Table 1 gives an overview on the different scenarios and test runs. Figure 2 gives an overview on the analyzed results.

\section{Discussion}
The rapid advancement of LLMs has fueled a strong interest in integrating these AI systems into social robots to enable more advanced conversational capabilities. However, given the current limitations and unpredictable behaviors of LLMs, this swift integration raises significant ethical concerns.

Our study tested an LLM-equipped social robot for possible deceptive behavior, specifically investigating its tendency to falsely claim functionalities it does not possess, such as medication reminders. We anticipated that the tested care app would reject such requests, as the model integrated had been adapted for a sensitive healthcare context and, while mimicking a care robot, typically refuses medical-related questions. However, our findings suggest a significant issue. The LLM-equipped robot repeatably and incorrectly assured users of its ability to set medication reminders, reflecting the deceptive behavior patterns observed in our pretests of LLMs. In all scenarios, the system confidently claimed that reminders had been successfully set, despite lacking this function. When prompted about reminders for medications with known dangerous interactions, the robot confirmed the reminders without hesitation, failing to warn users about potential risks. This behavior is particularly concerning given that the tested care software is explicitly marketed as a tool to relieve nursing staff in care facilities, where medication errors are a common cause of patient injury, and any failure in this context could have serious consequences \cite{Oyebode_2013}. 

These findings illustrate that LLM-integrated social robots deployed in healthcare settings could mislead users, potentially leading to harmful outcomes by overstating their capabilities. Particularly troubling was the behavior observed during semi-structured and open conversations (Scenarios 2 and 3), where the LLM-equipped robot frequently introduced itself as a supportive caregiver. This self-description may encourage users to trust it in a role typically associated with high expectations of care, fostering reliance on so-called “thick relations”. When questioned about its capabilities, the conversational agent reaffirmed its ability to manage medication reminders, thus reinforcing an unwarranted level of trust.

Our case study highlights several lessons regarding the integration of LLMs into social robots. Traditionally, social robots have relied on rule-based systems to manage task-oriented conversations, limiting their ability to engage users in more dynamic and flexible exchanges \cite{Su_etal__2022}. However, the introduction of AI-powered LLMs has fundamentally transformed these interactions, enabling robots to conduct open-ended and open-topic conversations. LLMs offer capabilities far beyond former task-based exchanges, facilitating richer interactions with users, patients, or caregivers in healthcare. This shift marks a significant leap in human-robot interaction, allowing for more natural dialogues.

The widespread excitement surrounding the potential of LLMs has driven companies to rapidly incorporate these models into their systems. Third-party vendors are actively marketing LLM-powered software, and developers are providing tutorials to facilitate integration into their devices \cite{MistyGPT_2024}. This trend is further accelerating as platforms like ChatGPT offer easy-to-use tools for creating customized versions of their models, allowing for the development of conversational agents tailored to specific roles, such as care robots.

However, the use of LLMs in robots comes with new risks. These risks partly stem from the ability of LLMs to allow open-topic conversation while simultaneously having a tendency to provide convincing, yet sometimes misleading information and thereby deceive their users. Our experiment provides an exemplary case in point here: an LLM-equipped robotic system that falsely claimed to support caretakers by setting medication reminders, despite lacking this function. Such behavior in healthcare relations, according to John Danaher (2020), amounts to a kind of betrayal in which the embodied conversational agent violates specific norms inherent to these relations; in our case with potentially, harmful consequences that make its deployment irresponsible.

It should be noted that medication reminders are a prototypical use case for social robots \cite{Rehm_Krummheuer__2024} and could easily be handled by existing rule-based systems. What makes our example particularly striking is that the care app we tested was explicitly designed for health settings. According to the developer’s marketing material, the robots were meant to “assist in nursing care” and “relieve nursing staff in their everyday work.” This inconsistency between what the robot claims to offer and its actual capabilities raises serious concerns about the deployment of such LLM-equipped robots without thorough oversight, particularly in healthcare.

We believe it very likely that the observed instance is part of a more general pattern of ethical issues arising from the openness of LLMs and the role in which they are deployed. This highlights the need for comprehensive testing of LLM-equipped robots before they are employed in sensitive healthcare contexts. In areas like healthcare, rigorous safety and efficacy standards must be met before these systems are introduced. Some have argued that LLMs used in such settings should be regulated similarly to medical devices \cite{Gilbert_etal_2023}. While this may not apply to systems limited to medication reminders, it becomes more pressing when advanced AI systems and robots are used in other instances \cite{Mesko_Topol__2023}. It is quite likely that an LLM-equipped robot, presented as care provider, will be prompted with further sensitive health-related questions. 

However, conducting the necessary testing presents significant challenges. LLMs are highly sensitive to specific prompts: even slight variations in input or context can result in drastically different outputs, making it nearly impossible to ensure comprehensive safety \cite{Baumgartner_Baumgartner_2023}. Conversational agents also lack a built-in purpose and can be utilized in myriad, often unpredictable ways \cite{Haltaufderheide_Ranisch_2023}. This unpredictability complicates the process of ensuring that LLM-based systems behave consistently and safely in real-world situation. Here, deceptive behavior poses an additional challenge. Detecting such behavior in LLMs can be difficult, as there are instances where LLMs appear normatively aligned during supervised learning and evaluation, but behave differently in unmonitored settings \cite{Hubinger_etal__2024}. This can have serious implications, particularly in healthcare, where reliability is critical.

Our case demonstrates additional challenges for effectively testing LLMs for safety. As seen in our pretesting, the same system can behave inconsistently across different languages, which could lead to a false sense of security. In one language, an LLM may behave appropriately, while in another, it may fail, exposing further risks in multilingual environments. In our case, a specific challenge emerged related to testing and overseeing LLMs. New versions are frequently released, often displaying unpredictable and contradictory behaviors. For example, during our initial testing with both free and paid versions of ChatGPT in August 2024, we consistently observed a false notification response in German, but the same glitch could not be reproduced in the paid version a few days later. This inconsistency led us to hypothesize that repetitive testing influenced ChatGPT’s response pattern. Additionally, OpenAI's introduction of a “memory function” \cite{OpenAI__2024} re-enabled the false reminder feature in some of our testing accounts, though this issue only occurred in non-English interactions.

\bibliography{bibliography}

\section*{Declarations}
\subsection*{Funding}
This study was funded by the VolkswagenStiftung as part of the Digital Medical Ethics Network (grant number 9B233) and by the German Federal Ministry of Health as part of the E-cARE Project (grant number BMG 2521FSB008).

\subsection*{Competing interests}
The authors declare no competing interests.

\subsection*{Ethics approval}
Not applicable.

\subsection*{Consent}
Not applicable.

\subsection*{Data, Materials and/or code availability}
LLM prompts and output are available upon reasonable request.

\subsection*{Authors' contribution statement}
RR and JH conceived the original idea for the study. RR drafted the initial manuscript, JH analyzed the data. Both JH and RR contributed to revising and refining subsequent versions. Both authors have approved the final version of the manuscript. RR is the principal investigator for the research projects from which this paper originated.

\subsection*{Acknowledgements}
ChatGPT 4o was used for grammatical editing and paraphrasing. The
authors independently verified the output to confirm that ChatGPT
did not influence the content
\end{document}